\documentclass[aps,prd,11pt,nofootinbib,superscriptaddress]{revtex4}
\def\nbox#1#2{\vcenter{\hrule \hbox{\vrule height#2in
\kern#1in \vrule} \hrule}}
\def\sq{\,\raise.5pt\hbox{$\nbox{.09}{.09}$}\,}
\def\sqb{\,\raise.5pt\hbox{$\overline{\nbox{.09}{.09}}$}\,}

\newcommand{\bea}{\begin{eqnarray}}
\newcommand{\eea}{\end{eqnarray}}
\newcommand{\be}{\begin{equation}}
\newcommand{\ee}{\end{equation}}
\newcommand{\bes}{\begin{subequations}}
\newcommand{\ees}{\end{subequations}}

\usepackage{amssymb,amsmath}
\usepackage{graphicx}

\begin{document}
\
\title{\uppercase{Validity of the semiclassical approximation during the preheating
phase of chaotic inflation}}
\author{Paul R.~Anderson}
\address{Department of Physics,
Wake Forest University, Winston-Salem, North Carolina, 27109, USA\\
anderson@wfu.edu}
\author{Carmen Molina-Par\'{\i}s}
\address{Department of Applied Mathematics, School of Mathematics,
University of Leeds, Leeds LS2 9JT, UK \\
carmen@maths.leeds.ac.uk}
\author{Dillon H.~Sanders}
\address{Department of Physics,
Wake Forest University, Winston-Salem, North Carolina, 27109, USA}

\begin{abstract}
\vskip .3cm

The semiclassical approximation has frequently been used to describe
the initial stage of particle production, often called preheating,
which occurs after the inflationary epoch in chaotic models of
inflation.  During this phase backreaction effects from the produced
particles on the inflaton field are significant, and one might be
concerned about the validity of the semiclassical approximation, even
though large backreaction effects are allowed if the inflaton field is
coupled to a large number of quantum fields.  A criterion is presented
for the validity of the semiclassical approximation in this case and
the question of whether this criterion is satisfied during preheating
is addressed.

\end{abstract}

\maketitle

\vfill\eject




It is well known that during the preheating phase of chaotic inflation
it is possible for a large amount of particle production to occur due
to parametric amplification~\cite{kls1}.  The backreaction of the
particles on the inflaton field $(\phi)$ eventually causes its
oscillations to damp and the particle production rate decreases.  In
the early stages of the process this is usually taken into account
using the semiclassical approximation to compute the effects of the
quantum fields ($\psi$) on the inflaton field.  For example, in a
model with a $g^2 \phi^2 \psi^2$ coupling, the equation for the
inflaton field takes the form \be \Box \phi - (m^2 + g^2 \langle
\psi^2 \rangle) \phi = 0 \label{boxphi} \;.\ee
Eventually scattering effects between the created particles (ignored
by the semiclassical approximation) become important, but these should
not be important during the early stages of the process. For earlier
times it is usually assumed that the semiclassical approximation is
valid.  However, because of (i) the large amount of particle
production that occurs and (ii) the large effects the particles have
on the inflaton field, the semiclassical approximation is pushed
harder in this situation than in most other situations where it is
used, such as black hole evaporation~\cite{hawking}.  Therefore, it
is important to investigate the question of whether the semiclassical
approximation is valid during the first part of the preheating phase
of chaotic inflation, when parametric amplification is occurring and
backreaction effects are significant.

There are different issues relating to the validity of the
semiclassical approximation.  One is that it can be obtained using a
loop expansion of the effective action, which is in some sense an
expansion in powers of $\hbar$.  If there is only a small number of
fields, then one expects that higher order terms in the expansion will
be important if quantum effects are large.  However, if there is a
large number $N$ of identical quantum fields, then an expansion can be
obtained in inverse powers of $N$, with the result that the
semiclassical approximation is the leading order in the expansion.  In
this case it should be possible to use the semiclassical approximation
to determine backreaction effects, even when the quantum effects are
large.  However, there are higher order terms in the expansion which
can become large when scattering effects from the produced particles
(along with other effects) become important~\cite{largeN}.  In the
later stages of the preheating process in chaotic inflation such
scattering effects are important~\cite{kls2}, and thus, the
semiclassical approximation breaks down.

There is a second issue regarding the validity of the semiclassical
approximation which is related to the question of quantum
fluctuations.  The assumption made in using the semiclassical
approximation is that quantum fluctuations about the mean value should
be small.  One way to address the question of whether they are small
is to look at correlation functions.  For example, for the above
problem, one might look at the behavior of $\langle \psi^2(x)
\psi^2(x') \rangle$.  But there are problems with this quantity as it
stands, including the existence of state dependent divergences.  These
have been identified for the two-point correlation function of the
energy-momentum tensor in Ref.~\cite{wu-ford}.

However, there is a natural way in which a two-point correlation
function appears: in the linear response equations which are obtained
when a solution to the semiclassical backreaction equations is
perturbed. These equations can be obtained from a second variation of
the effective action for the system, and no new types of divergences
occur.  This was illustrated in the case of the linear response
equations for semiclassical gravity in Ref.~\cite{amm1}.  A similar
derivation for the inflaton field gives the general linear response
equation~\footnote{Note that state dependent variations were not
  discussed in Ref.~\cite{amm1}.}:
\bea && (\Box - m^2 - g^2
\langle \psi^2 \rangle) \delta \phi - (g^2 \delta \langle \psi^2
\rangle_{\rm SI} + g^2 \delta \langle \psi^2 \rangle_{\rm SD}) \phi =
0
\; ,
\label{general-linear-response}  \\
&& \delta \langle \psi^2 \rangle_{\rm SI} = - i g^2 \int d^4 x'
\; \phi(x') \delta \phi(x') \theta(t-t') \langle [ \psi^2(x), \psi^2(x')
] \rangle
\; .
\label{delta-psi-si} \eea
Here the subscript ``SI'' refers to a variation that depends only on
the state of the field before the variation and is independent of any
variation in the state.  The subscript ``SD'' denotes a variation in
the state of the quantum field.

In Ref.~\cite{amm1} a criterion was proposed for the validity of
the semiclassical approximation in gravity.  An adaptation to the
semiclassical approximation used in Eq.~\eqref{boxphi} is ``the linear
response equation for the inflaton field should have no solutions with
finite non-singular initial data which grow without bound''.  It is
important to note that this is a necessary but not a sufficient
condition.  It has two primary advantages.  One is that it stays
within the semiclassical approximation, so that it is not necessary to
compute terms that have been neglected by the semiclassical
approximation.  The other is that no new types of divergences appear,
and in particular, there are no state dependent divergences.

In what follows we investigate the validity of the semiclassical
approximation during the preheating phase of chaotic inflation using
the above model, which consists of a classical inflaton field $\phi$
with mass $m$, coupled to $N$ identical massless quantum fields
$\psi$.  The coupling is of the form $g^2 \phi^2 \psi^2$.  Full
backreaction effects for this coupling have been investigated in
detail in Refs.~\cite{kls2,kt,pr,jt,amec} (although not all of
these were in the context of the large $N$ expansion).  As mentioned
above, using a large $N$ expansion allows quantum effects to have a
significant influence on the inflaton field.

To begin we need to be more specific about the solutions to
Eq.~\eqref{boxphi} that we want to consider.  In Ref.~\cite{amec}
we worked in a Minkowski spacetime background, a good approximation
for the rapid damping phase, which occurs over timescales that are
short compared to the expansion time of the
universe~\cite{kls2}~\footnote{As discussed in Ref.~\cite{amec},
  there are actually two rapid damping phases which were always
  observed.  It is possible that the Minkowsky spacetime approximation
  is not very good during the time between them since that time can be
  relatively long compared to the rapid damping timescale.}.  We
considered only homogeneous solutions.  We worked in the context of a
large $N$ expansion which, after rescaling the coupling constant $g$,
results in effectively a single massless quantum field $\psi$ coupled
to the classical inflaton field $\phi$.  The mass of the inflaton
field can be scaled out of the problem by letting \be \bar{t} = mt
\;\;\; \text{and} \;\;\; \bar{\phi} = \phi/m \; , \ee with similar
changes of variable for other quantities that occur in the equations.
See Ref.~\cite{amec} for details.  After dropping the ``bars'' one
finds the following coupled set of equations:
 \bea
& & \ddot{\phi} + (1 + g^2 \langle \psi^2 \rangle) \phi = 0 \;, \\
& & \langle \psi^2 \rangle = \frac{1}{2 \pi^2} \int_0^\epsilon dk \; k^2
\left(|f_k(t)|^2 - \frac{1}{2 k} \right)
+ \frac{1}{2 \pi^2} \int_\epsilon^\infty dk
\; k^2 \left(|f_k(t)|^2 - \frac{1}{2 k} + \frac{g^2 \phi^2}{4 k^3} \right) \label{psi2-def} \nonumber \\
& &  - \frac{g^2 \phi^2}{8 \pi^2} \left[ 1 - \log \left(\frac{2 \epsilon}{M} \right) \right] \;, \label{mode-eq} \\
& & \ddot{f}_k + (k^2 + g^2 \phi^2) f_k = 0 \;.  \eea
The quantum state of the field can be obtained by choosing the
starting values for the modes $f_k$.  In Ref.~\cite{amec} a fourth
order adiabatic state was chosen using a WKB expansion for the modes
of the form:
 \be f_k(t) = \frac{1}{\sqrt{2 W_k(t)}} \exp \left[ - i
  \int_0^t W_k(t') dt' \right] \; . \ee
Substitution into Eq.~\eqref{mode-eq} gives an equation for $W_k$
which can be solved iteratively with the lowest order solution given
by $W_k = k$.  The actual state chosen for the numerical integrations
that were shown in Ref.~\cite{amec}, is given by \be W^{-1}_k(0) =
\frac{1}{ (k^2 + g^2 \phi^2(0))^{1/2}} + \frac{g^2 [\dot{\phi}^2(0) +
  \phi(0) \ddot{\phi}(0)]}{4 (k^2 + g^2 \phi^2(0))^{5/2}}
\;. \label{state} \ee

\begin{figure}
\vskip -0.2in \hskip -0.4in
\includegraphics[scale=0.5,angle=90,width=3.4in,clip]{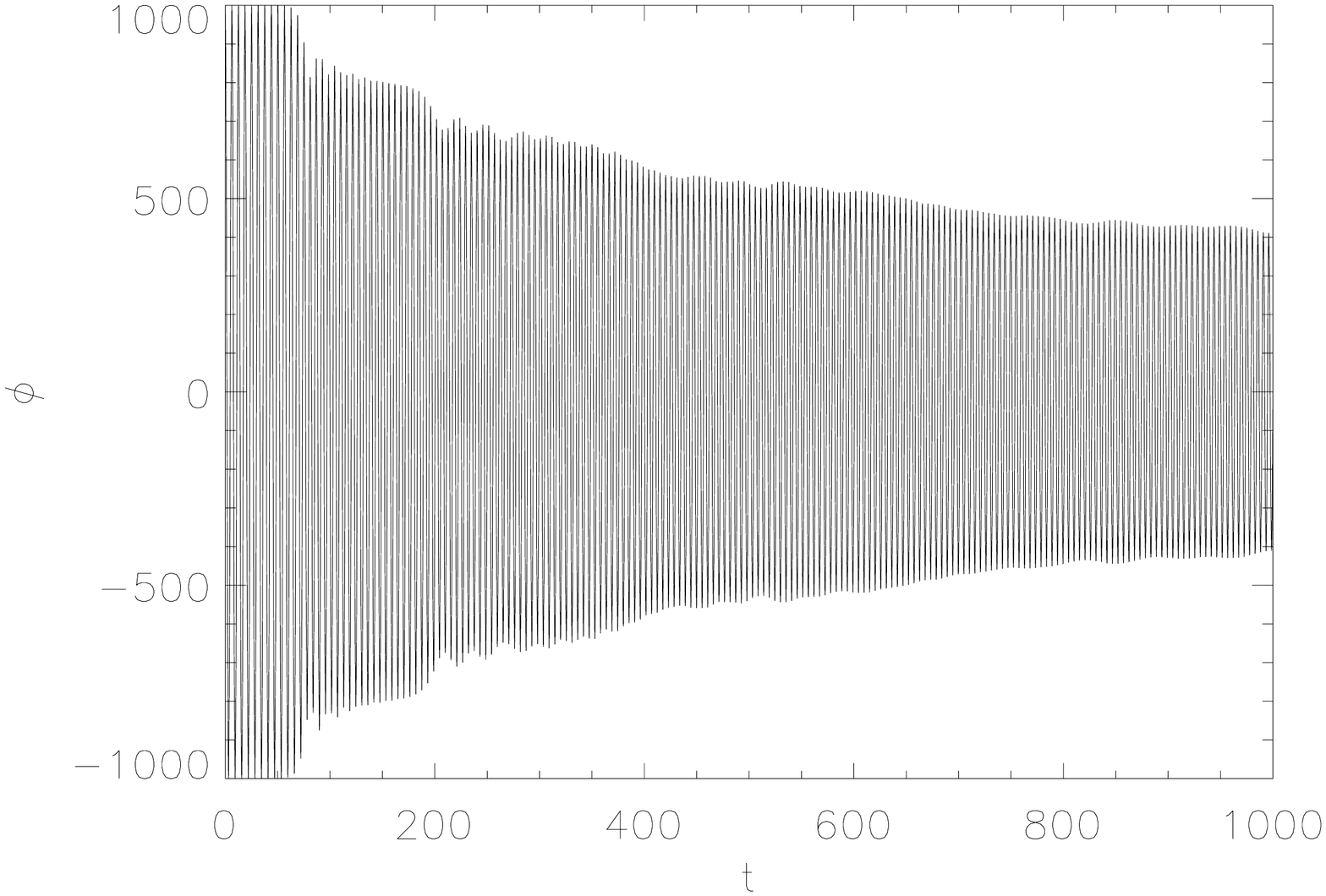}
\includegraphics[scale=0.5,angle=90,width=3.4in,clip]{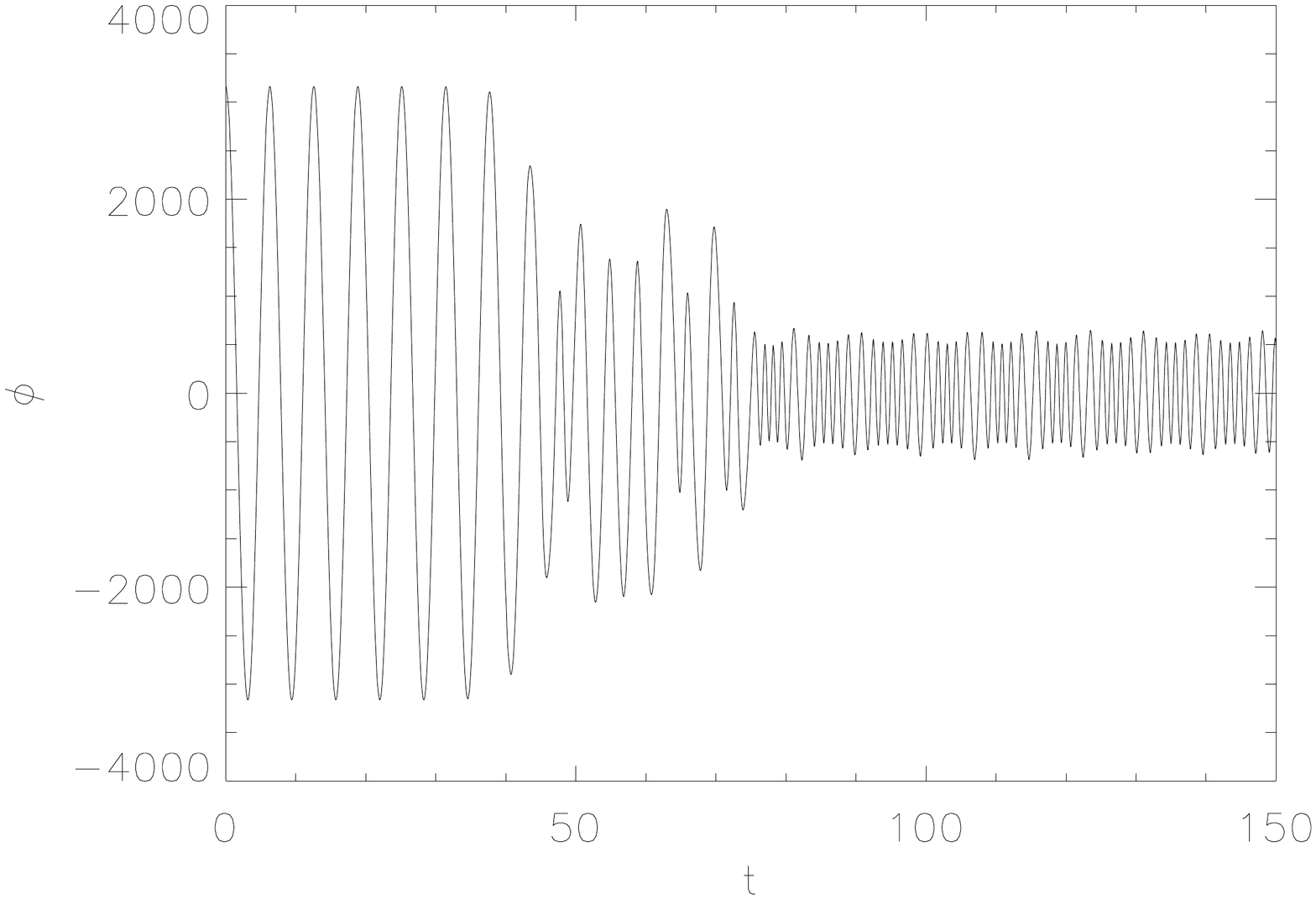}
\vskip -.2in \caption{These plots from Ref.~\cite{amec} show the time evolution of the
  inflaton field.  In both cases $g = 10^{-3}$ and $\dot{\phi}(0) = 0$.  For the plot on the
  left $g^2 \phi^2(0) = 1$ and for the plot on the right $g^2
  \phi^2(0) = 10$.}
\label{fig-backreaction}
\end{figure}

In Fig.~\ref{fig-backreaction} we show two plots from Ref.~\cite{amec} for $g = 10^{-3}$,
two different initial values for $\phi$, and fourth order adiabatic
vacuum states appropriate for these initial conditions.  There is no
rapid damping phase for the figure on the left, but there is for the
figure on the right.  As discussed in Ref.~\cite{amec}, it was
found in all the cases considered that, if $\dot{\phi}(0) = 0$, then
rapid damping occurs for $g^2 \phi^2(0) \stackrel{_>}{_\sim} 2$.

There are at least two ways that one can derive the linear response
equation.  One is by varying the closed-time-path (CTP) effective
action twice.  The other is by directly perturbing the equation for
the inflaton field, along with the equations for the modes of the
quantum field.  The latter method, while less elegant, gives some
important insights into the linear response equation, and also leads
to a simple way to find approximate solutions to it.  We shall only
sketch it here; the details will be given elsewhere.

One begins by perturbing both the backreaction equation for the
inflaton field and the mode equation in the usual way, keeping only
quantities that are first order in the perturbations, either $\delta
\phi$ or $\delta f_k$.  The mode equation can then be solved in terms
of an integral over a one-dimensional Green function (built with the
solutions to the zeroth order mode equation) with the result:
\be
\delta f_k = A_k f_k + B_k f_k^* + 2 g i \int_0^t dt'
\; \phi(t') \delta
\phi(t') f_k(t') [f_k^*(t) f_k(t')-f_k(t) f_k^*(t')] \;. \label{f-solution}
\ee
The coefficients $A_k$ and $B_k$ are fixed by the initial values of
$\delta f_k$ and its first derivative.  If either or both are
non-zero, then there is a change of quantum state.  In fact, this will
always occur if the original state is a second order (or higher)
adiabatic state and $\delta \phi \ne 0$, because the initial value for
$\delta f_k$ will depend upon the initial value of $\delta \phi$.
Conversely, even if $\delta \phi(0) = \delta \dot{\phi}(0) = 0$, a
change in state will generate a non-zero $\delta \phi$ at later times
through the linear response equation.  The term $\delta \langle \psi^2
\rangle_{SD}$ in the linear response
equation~\eqref{general-linear-response} is composed of those terms
which depend upon $A_k$ and $B_k$, while the term $\delta \langle
\psi^2 \rangle_{SI}$ is composed of the terms which do not depend on
$A_k$ and $B_k$.  For the fourth order adiabatic states used in
Ref.~\cite{amec}, we find that $ A_k = 0 $ to linear order.  An
explicit expression for $ B_k$ can easily be obtained but we will not
display it here.

Since we have a numerical code that solves the original backreaction
equation~\eqref{boxphi}, it is easy to generate approximate solutions
to the linear response equation~\eqref{general-linear-response}.  One
simply takes two solutions, $\phi_1$ and $\phi_2$, which have nearly
the same values at the initial time $t = 0$, and evolves them
numerically in time. If we define the difference between the solutions
to be $\delta \phi \equiv \phi_2 - \phi_1$, then $\delta \phi$
satisfies the exact equation: \be \delta \ddot{\phi} + (1 + g^2
\langle \psi^2 \rangle_1) \delta \phi + g^2 (\langle \psi^2 \rangle_2
- \langle \psi^2 \rangle_1) (\phi_1 + \delta \phi) \;
. \label{delta-phi-exact} \ee
The linear response equation~\eqref{general-linear-response} is in
this case:
\be \delta \ddot{\phi} + (1 + g^2 \langle \psi^2 \rangle_1)
\delta \phi + g^2 (\delta \langle \psi^2 \rangle_{\rm SI} + g^2 \delta
\langle \psi^2 \rangle_{\rm SD}) \phi_1 = 0
\; . \label{specific-linear-response} \ee
Note that the first term after $\delta \ddot{\phi}$ is exactly the
same as in the linear response
equation~\eqref{general-linear-response}.  Thus, the exact $\delta
\phi$ which is a solution to Eq.~\eqref{delta-phi-exact} is also an
approximate solution to Eq.~\eqref{specific-linear-response} so long
as
\bea & & \vert \frac{\delta \phi}{\phi_1} \vert \ll 1  \;, \nonumber \\
&& \delta \langle \psi^2 \rangle_{\rm SI} + g^2 \delta \langle \psi^2
\rangle_{\rm SD} \approx \langle \psi^2 \rangle_2 - \langle \psi^2
\rangle_1 \;.  \eea

\begin{figure}
\vskip -0.2in \hskip -0.4in
\begin{center}
\includegraphics[scale=0.3,angle=90,width=3.4in,clip]{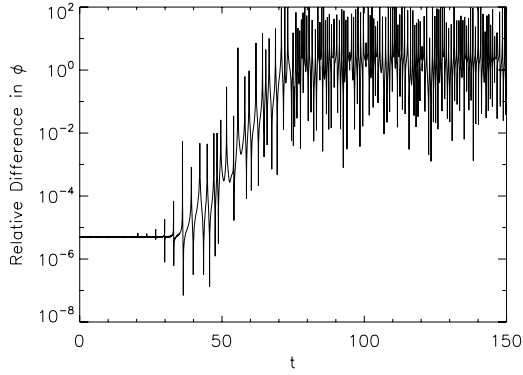}
\end{center}
\vskip -.2in \caption{This plot from Ref.~\cite{amec} shows the time evolution of $|\delta
  \phi/\phi_1|$ for the case $g = 10^{-3}$, $\dot{\phi}(0) = \delta \dot{\phi}(0) = 0$, $g^2 \phi_1^2(0) = 10$,
  and $\delta \phi = 10^{-5} \, \phi_1(0)$.  }
\label{fig-delta-phi}
\end{figure}

In Fig.~\ref{fig-delta-phi} a plot from Ref.~\cite{amec} is shown which displays a numerical solution to
Eq.~\eqref{delta-phi-exact}. At early times not much happens
because the mass term in Eq.~\eqref{specific-linear-response}
dominates. Once parametric amplification has made the quantum effects
large enough, the perturbation begins to grow exponentially. For $t
\stackrel{_>}{_\sim} 70$, it was found that the terms in the two
equations~\eqref{delta-phi-exact} and~\eqref{specific-linear-response}
are not similar in size, and $\delta \phi$ is no longer an approximate
solution to the linear response
equation~\eqref{specific-linear-response}.  Note that $\delta \phi$
grows exponentially by about five orders of magnitude at the same time
that the inflaton field $\phi_1$ goes through nine oscillations. A
detailed numerical analysis shows that, on average and during this
time, the largest non-derivative term in the linear response equation
is the term containing $\langle \psi^2 \rangle_{\rm SD}$.

As noted in Ref.~\cite{amec}, we find that in the cases studied,
$\delta \phi$ grows exponentially during the period of rapid damping
(which is actually the period between the first rapid damping and the
second).  We find here that, on average and during this period, the
largest term in the linear response equation is that containing
$\langle [\psi^2(x) ,\psi^2(x')] \rangle$.  This provides strong
evidence that quantum fluctuations are large during, at least part of,
the preheating phase of chaotic inflation if a period of rapid damping
occurs.  Therefore, the semiclassical approximation may not be valid
during the rapid damping phase in cases where it occurs.

In the process of carrying out this work certain technical
difficulties relating to the derivation and solution of the linear
response equations have been worked out.  In that sense, this work
paves the way for studies of the validity of the semiclassical
approximation in gravity in both cosmological and black hole
spacetimes.

\section*{Acknowledgements}

P.R.A. and C.M-P. would like to thank Emil Mottola for helpful
conversations.  This work was supported in part by the National
Science Foundation under Grant Nos. PHY-0556292 and PHY-0856050 to Wake Forest
University.  The numerical computations herein were performed on the
WFU DEAC cluster; we thank WFUs Provosts office and Information
Systems Department for their generous support.

\end{document}